\documentclass[review]{elsarticle}

\usepackage{
hyperref,amsmath,amssymb}

\journal{Journal of \LaTeX\ Templates}

\begin{document}

\begin{frontmatter}
\title{Application of Spin-Exchange Relaxation-Free Magnetometry to the Cosmic Axion Spin Precession Experiment}

\author[1]{Tao Wang\corref{mycorrespondingauthor1}}
\cortext[mycorrespondingauthor1]{Corresponding author}
\ead{taowang@berkeley.edu}
\author[4]{Derek F. Jackson Kimball}
\author[3]{Alexander O. Sushkov}
\author[3]{Deniz Aybas}
\author[2]{John W. Blanchard}
\author[2]{Gary Centers}
\author[1]{Sean R. O' Kelley}
\author[2]{Arne Wickenbrock}
\author[5]{Jiancheng Fang}
\author[2,1,6]{Dmitry Budker\corref{mycorrespondingauthor2}}
\cortext[mycorrespondingauthor2]{Corresponding author}
\ead{budker@uni-mainz.de}

\address[1]{Department of Physics, University of California, Berkeley, California 94720-7300, USA}
\address[4]{Department of Physics, California State University, East Bay, Hayward, California 94542-3084, USA}
\address[3]{Department of Physics, Boston University, Boston, Massachusetts 02215, USA}
\address[2]{Helmholtz Institute Mainz, Johannes Gutenberg University, 55099 Mainz, Germany}
\address[5]{School of Instrumentation Science and Opto-Electronics Engineering, Beihang University, Beijing 100191, PRC}
\address[6]{Nuclear Science Division, Lawrence Berkeley National Laboratory, Berkeley, California 94720, USA}

\begin{abstract}
The Cosmic Axion Spin Precession Experiment (CASPEr) seeks to measure oscillating torques on nuclear spins caused by axion or axion-like-particle (ALP) dark matter via nuclear magnetic resonance (NMR) techniques. A sample spin-polarized along a leading magnetic field experiences a resonance when the Larmor frequency matches the axion/ALP Compton frequency, generating precessing transverse nuclear magnetization. Here we demonstrate a Spin-Exchange Relaxation-Free (SERF) magnetometer with sensitivity $\approx 1~{\rm fT/\sqrt{Hz}}$ and an effective sensing volume of 0.1 $\rm{cm^3}$ that may be useful for NMR detection in CASPEr. A potential drawback of SERF-magnetometer-based NMR detection is the SERF's limited dynamic range. Use of a magnetic flux transformer to suppress the leading magnetic field is considered as a potential method to expand the SERF's dynamic range in order to probe higher axion/ALP Compton frequencies.
\end{abstract}
\end{frontmatter}
\section{\label{sec:level1}Introduction}

Dark matter and dark energy are the most abundant yet mysterious substances in the Universe. Axions and axion-like particles (ALPs; we do not distinguish between axions and ALPs in the following) have emerged as theoretically well-motivated dark-matter candidates \cite{graham2015experimental,khriplovich2012cp,asztalos2010squid,arik2009probing,graham2011axion,sikivie2014proposal,1367-2630-17-11-113025}. The Cosmic Axion Spin Precession Experiment (CASPEr) experiment searches for a time-varying axion field by using Nuclear Magnetic Resonance (NMR) techniques \cite{budker2014proposal,graham2013new,blanchard2015measurement,Ledbetter2013}. CAPSEr is projected to realize a sensitivity to axions and ALPs beyond the current astrophysical and laboratory limits \cite{graham2013new}.

As discussed in \cite{budker2014proposal,graham2013new}, a dark-matter ALP field can cause oscillating torques on nuclear spins either by generating an oscillating nuclear electric dipole moment (EDM) that interacts with a static electric field or through an oscillating ``ALP wind'' that acts as a pseudo-magnetic field along the relative velocity vector between the sample and the dark matter. The oscillation frequency of the torque is given by the ALP Compton frequency $\omega_a$. In CASPEr, a sample of nuclear spins is polarized along a leading magnetic field, and if the Larmor frequency matches $\omega_a$, a resonance occurs and precessing transverse magnetization is generated. The initial plan for CASPEr employs Superconducting Quantum Interference Device (SQUID) magnetometers to search frequencies $\lesssim 1~{\rm MHz}$ (roughly corresponding to an applied magnetic field below $0.1~{\rm T}$ depending on the sample), and inductive detection using an LC circuit for frequencies $\gtrsim 1~{\rm MHz}$.

Another possibility for NMR detection is the use of an optical atomic magnetometer \cite{budker2013optical}. In particular, a state-of-the-art Spin-Exchange Relaxation-Free (SERF) magnetometer has realized a sensitivity of 160 $\mathrm{aT/\sqrt{Hz}}$ in a gradiometer arrangement, and its quantum noise limit is 50 $\mathrm{aT/\sqrt{Hz}}$, which is the most sensitive magnetometer in the low-frequency region \cite{dang2010ultra}. This motivates consideration of SERF magnetometers for NMR detection in CASPEr \cite{budker2014proposal}. SERF magnetometers are applied in fundamental symmetry tests \cite{budker2013optical,brown2010new,romalis2014comment,chu2016search,ji2016searching}; they have better sensitivity than Superconducting Quantum Interference Devices (SQUIDs) in the low-field regime \cite{dang2010ultra,nenonen1996thermal}, which could in principle improve the sensitivity of the search for axion dark matter, but SERF magnetometers have a disadvantage of a smaller bandwidth than SQUIDs \cite{kominis2003subfemtotesla,drung19955}. With a magnetically shielded room, a SQUID magnetometer operated inside a LOw Intrinsic NOise Dewar (LINOD) could reach a noise level of about 260 $\mathrm{aT/\sqrt{Hz}}$ below 100 Hz, and achieve a noise level of 150 $\mathrm{aT/\sqrt{Hz}}$ between 20 kHz and 2.5 MHz \cite{storm2017ultra}. SERF magnetometers have demonstrated comparable magnetic-field sensitivities to those of SQUID magnetometers; however, they have certain advantages that may be important in specific applications. First and foremost, SERF magnetometers do not require cryogenics, they generally have the “1/f knee” at lower frequencies, and they are robust with respect to electromagnetic transients. There are also disadvantages such as generally lower dynamic range, bandwidth, the necessity to heat the sensor cells, larger sensor size, and the absence of the elegant gradiometric arrangements possible with SQUIDs. 

In Sec. II, CASPEr is summarized and the corresponding estimates for the axion induced signal shown. We then explore the potential of SERF magnetometry in CASPEr where an experimental arrangement is proposed and various sources of noise are considered. In Sec. III, we introduce a modification to the quantum noise equations to account for position dependent atomic absorption by the pump beam. We then present a 1 fT/$\sqrt{\mathrm{Hz}}$ magnetometer and demonstrate a measurement of the modified noise described by the equations.  A possible technique to significantly expand the bandwidth of the SERF axion search is also explored.

\section{SERF Magnetometers for Spin Precession Detection in CASPEr}

The CASPEr research program encompasses experiments employing established technology to search for an oscillating nuclear electric dipole moment (EDM) induced by axions or ALPs (CASPEr Electric) and search for direct interaction of nuclear spins with an oscillating axion/ALP field (axion wind; CASPEr Wind). The CASPEr-Wind and the CASPEr-Electric experiments have a lot of features in common. The proposal to use a SERF magnetometer for detection of spin precession may be applicable to both CASPEr-Wind and CASPEr-Electric although in the following we focus on CASPEr-Electric. The axion field can be treated as a fictitious AC-magnetic field acting on nuclear spins in an electrically polarized material \cite{budker2014proposal}
\begin{equation}
B_a(t)=\frac{\epsilon_S E^*d_n}{\mu}\textrm{sin}\left(\omega_a t\right)
\label{Eq_axion_field},
\end{equation}
where $\epsilon_S$ is the Schiff factor \cite{khriplovich2012cp}, ${E^*}$ is the effective static electric field acting on the atoms containing the nuclear spins of interest, $\mu$ is the nuclear magnetic moment, $\omega_a=m_a/\hbar$ is the frequency of the axion (we set c=1 in the paper), and $m_a$ is the mass of the axion. Note that the field oscillates at the Compton frequency of the axion. The nuclear electric dipole moment ($d_n$) generated by the axion dark matter can be written as \cite{graham2013new}
\begin{equation}
d_n \approx (10^{-25} e\cdot\textrm{cm})(\frac{\textrm{eV}}{m_a})(g_d \times \textrm{GeV}^2)
\label{Eq_nEDM}.
\end{equation}
where $g_d$ is the EDM coupling.

\begin{figure}
\centering
\includegraphics[width=6.5cm]{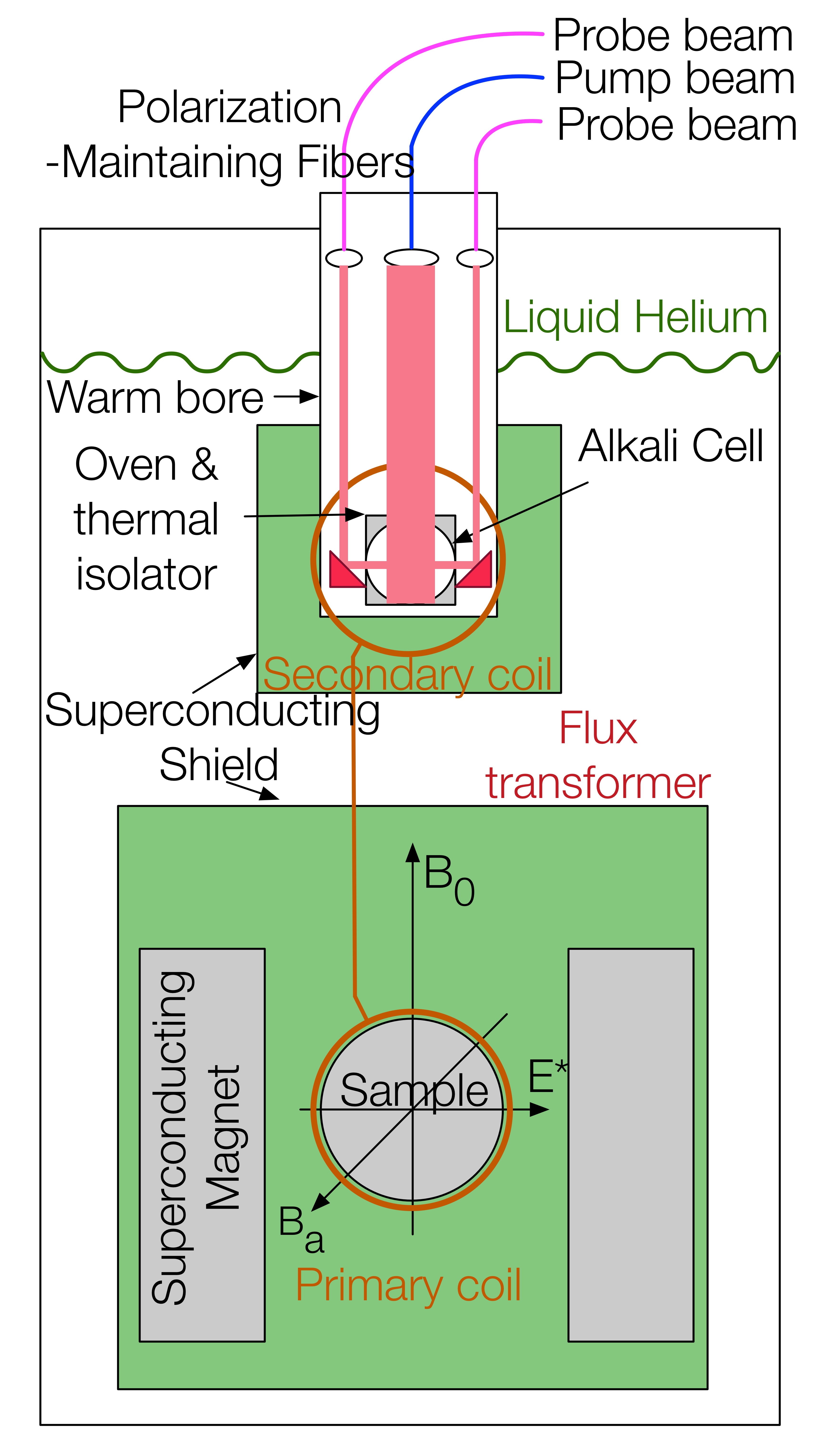}
\caption{\label{Fig_setup_casper-e} Conceptual schematic of the CASPEr experiment with a SERF magnetometer.}
\end{figure}

In the CASPEr experiment, the nuclear spins in a solid sample are prepolarized by either a several tesla magnetic field generated by superconducting coils or optical polarization via transient paramagnetic centers. The experiment is then carried out in a leading magnetic field ${B_{0}}$; the effective electric field ${E^*}$ inside the sample is perpendicular to ${B_{0}}$ as shown in Fig. \ref{Fig_setup_casper-e}. The time-varying moments induced by axion dark matter are collinear with nuclear spin. In the rotating frame, if there is a nucleon electric dipole moment, the nuclear spins will precess around the electric field, and this will induce a transverse magnetization, which can be measured with a sensitive magnetometer. The first generation CASPEr-Electric experiment will most likely employ a ferroelectric sample containing Pb as the active element. As mentioned in \cite{budker2014proposal}, $\rm{^{207}Pb}$  (nuclear spin $I=1/2$) has a nonzero magnetic dipole moment, and has a large atomic number ($Z$), which means it has a large Schiff factor (since the effect produced by the Schiff moment increases faster than $Z^2$) \cite{flambaum2008electric,bouchard2008nuclear}. The transverse magnetization of the ferroelectric samples caused by the axion field can be written as \cite{budker2014proposal,abragam1961principles}
\begin{equation}
M_a(t) \approx n_{Pb}p\mu\gamma_{Pb} \frac{1/T_b}{(1/T_b)^2+(\omega_0-{m_a}/{\hbar})^2 }B_a(t)
\label{Eq_transvers_magnetization},
\end{equation}
where $n_{Pb}$ is the number density of nuclear spins of $\rm{^{207}Pb}$, $p$ is the spin polarization of $\rm{^{207}Pb}$, $\mu=0.584\mu_N$ is the nuclear magnetic moment of $^{207}$Pb, $\gamma_{Pb}$ is the gyromagnetic ratio of $^{207}$Pb, $\omega_0$ is the spin-precession frequency in the applied magnetic field, we define $T_b=min\{T_2,\tau_a\}$ as the ``signal bandwidth time", $T_2$ is the transverse relaxation time of the nuclear spins,  and $\tau_{a}= 10^6h/m_{a}$ is the axion coherence time \cite{graham2013new}, which varies from $4 \times 10^5$ to $4 \times 10^{-3}$ s over the range of the axion masses from $10^{-14}-10^{-6}$ eV.

CASPEr searches for axion dark matter corresponding to axions of different masses by sweeping the applied magnetic field from zero to several T or higher, which in turn scans the NMR resonance frequency and sets the axion Compton frequency to which the apparatus is sensitive. Much of the interesting parameter space corresponds to field values that exceed the magnetic field limit of the SERF magnetometer. The large DC field problem can be solved using a flux transformer as shown in Fig. (\ref{Fig_setup_casper-e}), which acts as a ``DC magnetic filter" reducing the static magnetic field to keep the alkali metal atoms in the SERF regime. The flux transformer only picks up the time-varying component of the magnetic flux through the enclosed area. 

A SERF magnetometer has a narrow bandwidth of a few Hz \cite{allred2002high}; by applying a constant magnetic field along the pump-beam direction, the SERF magnetometer can be tuned to resonate at a higher frequency, which increases the detectable frequency range of the SERF magnetometer up to 200 Hz \cite{kornack2007low} or higher. The LOngitudinal Detection scheme (LOD) discussed in \cite{1742-6596-718-4-042051,granwehr2002new} can, in principle, fully remedy the disadvantage of the SERF magnetometer's limited bandwidth, which is discussed in the Appendix.

The alkali cell of the SERF magnetometer is heated to 373 K- 473 K in order to increase the alkali vapor density to improve the sensitivity. However, the ferroelectric sample is cooled down to a low temperature to increase the longitudinal relaxation time and the spin polarization of the sample \cite{mukhamedjanov411226suggested,budker2014proposal}. Again a flux transformer \cite{savukov2009mri,savukov2014multi} is a potential solution to this problem where, as shown in Fig. \ref{Fig_setup_casper-e}, the SERF magnetometer can be placed in a warm bore of a superconducting system containing the transformer coils and magnetic shields.

The magnetic flux through the primary coil can be written as \cite{graham2013new,sushkov2009prospects}
\begin{equation}
\Phi_p= \mu_0 \mu_r g N_{p} M_a A_{p}
\label{Eq_flux_pickup},
\end{equation}
where $\mu_r$ is the relative permeability of the ferroelectric sample, $\mu_r \approx 1$ for $\mathrm{PbTiO_3}$, $g\approx 1$ is the geometric demagnetizing factor \cite{sushkov2009prospects}, $A_{p}$ is the cross-section of the cylindrical sample, and $N_{p}$ is the number of turns of the primary coil. 

The flux transformer has an enhancement factor $(k_{FT}=B_{s}/B_{p})$, where $B_s$ is the magnetic field in the secondary coil, $B_p$ is the magnetic field in the primary coil, which can be calculated as
\begin{equation}
\begin{aligned}
k_{FT}&=\frac{N_p A_pB_s}{\Phi_p}=\frac{N_pA_p}{\Phi_p}\frac{\mu_0N_s}{l_s}\frac{\Phi_p}{L_s+L_p}=\frac{\mu_0 N_s}{l_s} \frac{N_p A_p}{L_s+L_p}
\label{Eq_kft},
\end{aligned}
\end{equation}
where $N_{s}$ is the number of turns of the secondary coil, $l_s$ is the coil length of the secondary coil, $L_{p}$ and $L_{s}$ are the inductances of the primary and the secondary coil, respectively. Inductances of multi-turn long solenoid coil can be written as 
\begin{equation}
\begin{aligned}
	L_p\approx \frac{\mu_0 N_p^2 A_p}{l_p},
	L_s\approx \frac{\mu_0 N_s^2 A_s}{l_s},
	\label{Eq_inductance}
	\end{aligned}
\end{equation}
where $A_{s}$ is the winding cross-section of the secondary coil, $l_p$ is the coil length of the primary coil. Inserting Eq.(\ref{Eq_inductance}) into Eq.(\ref{Eq_kft}), we have 
\begin{equation}
\begin{aligned}
 	k_{FT}&=\frac{1}{\frac{l_s}{l_p}\frac{N_p}{N_s}+\frac{A_s}{A_p}\frac{N_s}{N_p}};
 	\end{aligned}
 	\label{Eq_opt_kft}
\end{equation}
with $N_p/N_s=\sqrt{A_sl_p/A_pl_s}$, we have $L_p=L_s$, and the gain factor of Eq.(\ref{Eq_opt_kft}) has a maximum $\sqrt{A_pl_p/4A_sl_s}$. Careful consideration of the relative coil geometries, taking into account sample size and insulation requirements for example, is required to ensure this remains an enhancement factor. \footnote{For small samples and larger amounts of insulation between the flux transformer and the SERF magnetometer, the geometry may lead to an unfavorable $k_{FT}<1$. For example, a 1.6 cm diameter sample with a 6 cm diameter secondary coil, $l_p$ = 1.6 cm and $l_s$ = 2 cm, yields a $k_{FT}$ = 0.1, reducing the advantage of using a flux transformer. But for larger samples, the $k_{FT}$ grows greater than 1, making this approach especially useful.}

At a finite temperature, the flux transformer induces Johnson noise. The noise of the flux transformer and the SERF magnetometer system is determined by the noise of the flux transformer ($\delta B_{FT}$), the field enhancement coefficient, and the sensitivity of the SERF magnetometer ($\delta B_{SERF}$)
\begin{equation}
	\delta B_n=\sqrt{\delta B_{FT}^2+(\frac{\delta B_{SERF}}{k_{FT}})^2}.
\end{equation}

In this experiment, the flux transformer is made of zero-dissipation material, such as superconducting niobium or niobium-titanium wire and cooled with liquid helium to realize superconductivity \cite{clarke2004squid}. Type I superconductors are the ideal choice since the magnetic field cannot penetrate, making the Johnson noise negligible. However, the sweeping magnetic field reaches the critical field of these materials around 100 mT ($\approx 6$ MHz for Pb). Depending on the parameter space to be explored, the material should be chosen accordingly.

If $t<\tau_a$, the experimental sensitivity after measurement time $t$ can be written as \cite{budker2014proposal} 
\begin{equation}
\begin{aligned}
\frac{\Phi_{p}}{N_p A_p}=\frac{\delta B_n}{\sqrt{t}}.
\end{aligned}
\label{Eq_SERF_sensitivity}
\end{equation}
When $t>\tau_a$, and the experimental sensitivity after measurement time $t$ can be written as \cite{budker2014proposal} 
\begin{equation}
\begin{aligned}
\frac{\Phi_{p}}{N_p A_p}=\frac{\delta B_n}{(\tau_a t)^{1/4}}.
\end{aligned}
\label{Eq_SERF_sensitivity_2}
\end{equation}

When $t<\tau_a$ and $T_2<\tau_a$, the axion-nucleon coupling constant can be calculated by Eqs. (\ref{Eq_axion_field}-\ref{Eq_flux_pickup}) and (\ref{Eq_SERF_sensitivity}). When $\omega_m=m_a/\hbar$, the transverse magnetization is enhanced at the resonant point; we find
\begin{equation}
g_{d}[{\mathrm{GeV^{-2}}}]=\frac{4.5\times 10^{45}}{\mathrm{C\cdot m}}\frac{\delta B_{n} \times m_a[\textrm{eV}]}{ \mu_0 n_{Pb}p\gamma_{Pb} T_2 \epsilon_S E^*\sqrt{t}}
\label{Eq_gd}.
\end{equation}
When $t>\tau_a$ and $T_2<\tau_a$, the axion-nucleon coupling constant can be calculated by Eqs. (\ref{Eq_axion_field}-\ref{Eq_flux_pickup}) and (\ref{Eq_SERF_sensitivity_2}) as
\begin{equation}
g_{d}[{\mathrm{GeV^{-2}}}]=\frac{4.5\times 10^{45}}{\mathrm{C\cdot m}}\frac{\delta B_{n} \times m_a^{5/4}[\textrm{eV}]}{ \mu_0  n_{Pb}p\gamma_{Pb} T_2 \epsilon_S E^*(10^6h[\mathrm{eV\cdot s}] t)^{1/4}}
\label{Eq_gd_2}.
\end{equation}
When $T_2>\tau_a$, then 
\begin{equation}
g_{d}[{\mathrm{GeV^{-2}}}]=\frac{4.5\times 10^{45}}{\mathrm{C\cdot m}}\frac{\delta B_{n} \times m_a^{9/4}[\textrm{eV}]}{ \mu_0  n_{Pb}p\gamma_{Pb} \epsilon_S E^*(10^6h[\mathrm{eV\cdot s}])^{5/4}(t)^{1/4}}
\label{Eq_gd_3}.
\end{equation}

\begin{figure*}
\centering
\includegraphics[width=10cm]{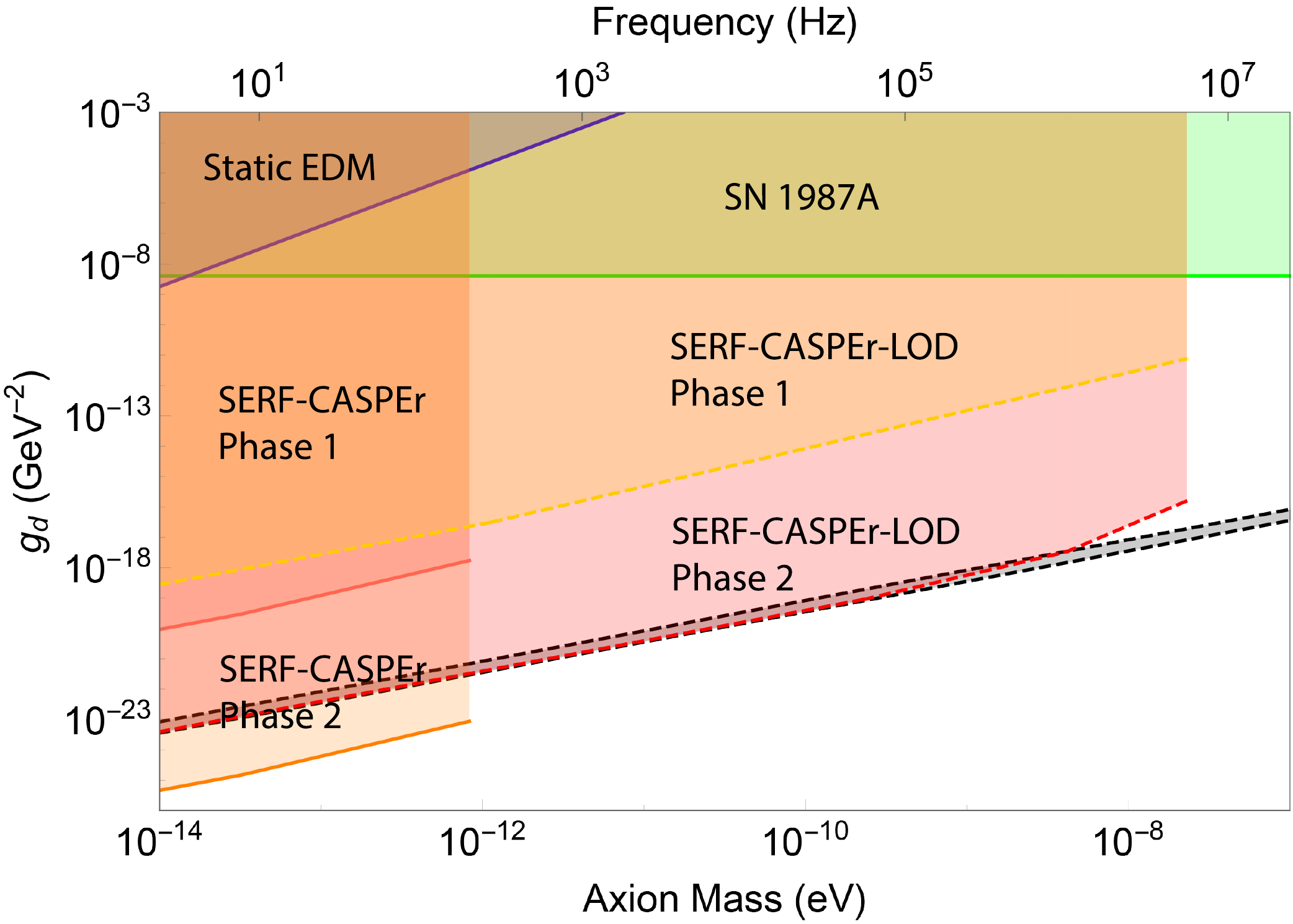}
\caption{\label{Fig_Mag2CASPEr} Sensitivity projection plot for axions. The blue region is excluded by static electric dipole moment experiments \cite{graham2013new}. Reference \cite{abel2017search} demonstrates a better static-EDM experimental results, which is in the sub-mHz range; out of the frequency range of the CASPEr experiment. The green region is excluded by the absence of excess cooling in supernova 1987A \cite{graham2013new}. The pink region is the Phase 1 detectable range of the SERF-CASPEr based on Eqs. (\ref{Eq_gd}) and (\ref{Eq_gd_2}). The orange region is the Phase 2 detecable range of the SERF-CASPEr based on Eqs. (\ref{Eq_gd}) and (\ref{Eq_gd_2}). The SERF-CASPEr-LOD sensitivity estimates are analagously shown in the Appendix using Eqs. (\ref{Eq_gd}-\ref{Eq_gd_3}) for Phase 1 and 2. The dashed black band is the quantum chromodynamics (QCD) axion \cite{graham2013new,diCortona2016}, reference \cite{guo2015electric} gives a larger result, which results in a sensitivity improvement approximately by a factor of 2.}
\end{figure*}

Here we assumed that the transverse relaxation time $T_2$ equals 5 ms in Phase 1 of the CASPEr experiment\cite{budker2014proposal}. The sample is paramagnetic purified $\rm{PbTiO_3}$, which is polarized by 20 T magnetic field at 4.2 K, yielding $p$ = 0.001 \cite{budker2014proposal}. The $E^*$ is assumed to be $3 \times 10^{10}$ V/m \cite{mukhamedjanov411226suggested}. We assumed $A_p = 78$ $\rm{cm^2}$, $A_s = 28$ $\rm{cm^2}$, $l_p$ = 10 cm, $l_s$ = 2 cm, and enhancement factor $k_{FT} \approx 2$, and the sensitivity of the SERF magnetometer is 50 $\rm{aT/\sqrt{Hz}}$ \cite{dang2010ultra}. In Phase 2, we assumed $T_2 = 1$ s, and $p$ = 1 \cite{budker2014proposal}. Utilizing the narrow bandwidth range, the measurement time of a single frequency point is assumed to be 36 hours. It will take approximately 1 year of continuous data acquisition to sweep the 200 Hz of parameter space. The detectable region for the SERF magnetometer is calculated with Eq.(\ref{Eq_gd}), which is plotted as the orange region in Fig. \ref{Fig_Mag2CASPEr}. From the figure, we can see that CASPEr experiments employing a SERF magnetometer for NMR detection can realize a sensitivity of $10^{-25}$ $\rm{GeV^{-2}}$. Technical noise such as that due to vibrations of the apparatus is a major concern for such experiments, but we point out here that the Q-factor of the axion oscillation is $\approx 10^6$, which is usually much higher than the Q-factor of the vibrational noises facilitating suppression of the latter.

\section{Model of SERF magnetometer noise limitations}

A SERF magnetometer is an alkali vapor atomic magnetometer, that works in the regime where the spin-exchange rate far exceeds the frequency of Larmor procession. In this regime spin-exchange relaxation is suppressed \cite{happer1973spin}. A circularly polarized pump beam is used to spin polarize the alkali atoms while a linearly polarized probe beam propagates perpendicularly through the cell. If a small magnetic field is applied perpendicular to the plane of the pump and probe beams, this will cause the spins to precess by a small angle and probe beam's plane of polarization to rotate by an angle proportional to the magnetic field due to the Faraday effect. The magnetic field can thus be determined by measuring the optical rotation angle.

The sensitivity with which spin precession can be measured determines the achievable sensitivity of the CASPEr experiment up to the point where the magnetization noise of the sample becomes dominant. To date, measurements of the sensitivity of the SERF magnetometers has been limited by Johnson noise from the magnetic shield, even with a low-noise ferrite magnetic shield \cite{kornack2007low,dang2010ultra}. The Low Intrinsic NOise Dewar (LINOD) reported in \cite{storm2017ultra} opens new possibilities for ultra-low magnetic noise superconducting shields, in which case the quantum noise and the technical noise could become the dominant noise sources in the future. Thus, studying the intrinsic noise of a SERF magnetometer , as we do in this work as a first step in the investigation of the possible application of SERF magnetometry to CASPEr, is essential for further improving sensitivity of the SERF magnetometry.

The SERF magnetometer has a fundamental sensitivity at the attotesla ($10^{-18}$ tesla) level \cite{budker2007optical,dang2010ultra,fang2014situ} limited by the spin-projection noise (SPN) and the photon shot noise (PSN) \cite{ledbetter2008spin}. In many practical implementations, the photon shot noise is a major contribution to the quantum noise limit for SERF magnetometers \cite{dang2010ultra}.

When photon shot noise is the dominant quantum noise source, the optimum sensitivity of a SERF magnetometer is achieved when the polarization of the atoms is 50\% \cite{kornack2005test,seltzer2008developments} and the power of the pump beam is chosen accordingly. Furthermore, a detuned pump beam causes light shifts, which can be treated as a fictitious magnetic field \cite{appelt1998theory,fang2014optimizations}. This light shift is conventionally eliminated by locking the pump beam's frequency to the resonance point. However, the on-resonance pump beam is strongly adsorbed by the non-fully polarized alkali atoms due to the larger optical depth. The absorption causes position-dependent polarization along the pump-beam propagation direction in the cell. A hybrid optical pumping scheme has been proposed to solve this problem \cite{romalis2010hybrid}, however to-date in practice the sensitivity of the hybrid SERF magnetometers have not surpassed the sensitivity of direct-optical-pumping-based potassium SERF magnetometers \cite{fang2014optimizations,ito2011sensitivity}.

Here we determine the noise limit for a direct-optical-pumping-based SERF magnetometer taking into account the absorption of the pump beam by the alkali atoms. Furthermore, the sensitivity of the SERF magnetometer is usually limited by the optical rotation measurement which we experimentally demonstrate along with the analytic absorption modification to the noise limit. 

The major sources of noise affecting SERF magnetometers can be divided into three categories, 1) Quantum noise (Spin-projection noise and Photon shot noise). 2) Technical noise (The probe beam polarization rotation noise caused by the Faraday modulator and Lock-in amplifier etc.). 3) Magnetic noise (Johnson noise of the magnetic shields).

The contribution to the apparent magnetic noise per root Hz measured by a SERF magnetometer associated with photon shot noise can be written as \cite{ledbetter2008spin}
\begin{equation}
\delta B_{PSN}=\frac{\hbar}{g_s \mu _B P_z \sqrt{nV}}\frac{2\sqrt{2}(R+\Gamma_{pr}+\Gamma_{SD})}{\sqrt{\Gamma_{pr}(OD)_0}}
\label{psn},
\end{equation}
where $\hbar$ is the reduced Planck constant, $g_s\approx2$ is the electron Lande factor, $g_s\mu_B/\hbar=\gamma_e$ is the gyromagnetic ratio of the electron, $\mu_B$ is the Bohr magneton, $P_z$ is the spin polarization along the pump beam, which is
\begin{equation}
P_z=\frac{R}{R+\Gamma_{pr}+\Gamma_{SD}}
\label{pol},
\end{equation}
 $n$ is the density of the alkali atoms, $V$ is the overlapping volume of the probe beam and the pump beam, $t$ is the measurement time, $R$ is the pumping rate of the pump beam, $\Gamma_{pr}$ is the pumping rate of the probe beam, $\Gamma_{SD}$ is the spin-relaxation rate caused by the spin destruction, and $OD_0$ is the optical depth on resonance. 

The apparent magnetic noise per root Hz measured by a SERF magnetometer associated with spin-projection noise can be written as \cite{ledbetter2008spin,seltzer2008developments}

\begin{equation}
\delta B_{SPN}=\frac{2\hbar\sqrt{(R+\Gamma_{pr}+\Gamma_{SD})}}{g_s \mu _B P_z \sqrt{nV}}
\label{spn}.
\end{equation}
Calculating the quadrature sum of Eq.(\ref{psn}) and Eq.(\ref{spn}), we find the expression for the total quantum noise
\begin{equation}
\begin{aligned}
\delta B_{qt}&=\frac{2\hbar(\sqrt{R+\Gamma_{pr}+\Gamma_{SD}})^{3}}{g_s \mu_B R \sqrt{n V}}\sqrt{1+\frac{2(R+\Gamma_{pr}+\Gamma_{SD})}{\Gamma_{pr}(OD)_0}}
\label{fn}.
\end{aligned}
\end{equation}
The sensitivity of the optical rotation measurement plays an important role in the atomic spin measurement, which can also be a limitation for the sensitivity of the SERF magnetometer. The optical rotation angle $\theta$ of a linearly polarized probe beam can be described as \cite{seltzer2008developments,quan2016far}
\begin{equation}
\begin{aligned}
\theta&=\frac{\pi}{2}nlr_ecP_xf_{D2}\frac{\Gamma_L/2\pi}{(\nu-\nu_{D2})^2+(\Gamma_L/2)^2}
\label{Eq_opticalrotation},
\end{aligned}
\end{equation}
where $l$ is the length of the optical path of the probe beam through the alkali cell, $r_e$ is the classical radius of the electron, $c$ is the speed of light, $P_x$ is the spin polarization projection along the X-axis, $f_{D2}$ is the oscillator strengths of the D2 line, $\Gamma_L$ is the pressure broadening caused by the buffer gas and quenching gas, $\nu$ is the frequency of the probe beam, and $\nu_{D2}$ is the resonance frequency of the D2 line.

Under conditions where the residual magnetic fields are well-compensated, a small magnetic field $B_y$ applied along the Y-axis causes the net spin polarization to precess, generating a non-zero spin projection along the X-axis, given by
\begin{equation}
P_x=\frac{\gamma_e R B_y}{(R+\Gamma_{pr}+\Gamma_{SD})^2+(\gamma_e B_y)^2+(\gamma_e B_{LS})^2}
\label{eq6},
\end{equation}
where $B_{LS}$ is the light shift. Such a Y-directed field can be used to calibrate a SERF magnetometer. If the calibration magnetic field applied along the Y-direction $B_y \ll (R+\Gamma_{pr}+\Gamma_{SD})/\gamma_e$, and the light shift is negligible, then Eq.(\ref{eq6}) can be simplified to
\begin{equation}
P_x=\frac{\gamma_e R B_y}{(R+\Gamma_{pr}+\Gamma_{SD})^2}
\label{eq7}.
\end{equation}

Combining the results of Eq.(\ref{eq7}) and Eq.(\ref{Eq_opticalrotation}), and assuming the wavelength of the probe beam is several hundreds GHz detuned (which depends on the pressure broadening $\Gamma_L$) to lower frequency from the D2 resonance frequency, we find an expression for optical-rotation-induced apparent magnetic noise

\begin{equation}
\delta B_m=\frac{4(R+\Gamma_{pr}+\Gamma_{SD})^2 [(\nu-\nu_{D2})^2+(\Gamma_L/2)^2] \delta \theta}{\gamma_e R nlr_ecf_{D2}\Gamma_L}
\label{probenoise},
\end{equation}
where $\delta \theta$ is the sensitivity of the optical rotation measurement in  $\rm{rad/\sqrt{Hz}}$.

To better describe the parameters determining the SERF noise limits, one must additionally account for the adsorption of the pump beam propagating through the cell, which is not linearly proportional to the power of the pump beam measured before the cell. The pump beam propagates along the cell with the pumping rate decreasing according to \cite{kornack2005test}

\begin{equation}
R(z)=R_{\rm{IN}}W \left[\frac{R_{\rm{IN}}}{\Gamma_{rel}} e^{\frac{R_{\rm{IN}}}{\Gamma_{rel}}-n\sigma(\nu)z}\right]
\label{absorption}
,
\end{equation}
where $R_{\rm{IN}}$ is the pumping rate of the pump beam entering the front of the cell. $R(z)$ is the pumping rate of the pump beam which propagates in the cell with a distance of $z$. $W$ is the Lambert W-function, which is the inverse of the function $f(W)=We^W$, $\Gamma_{rel}=\Gamma_{SD}+\Gamma_{pr}$. In a vapor with a large optical depth, the pumping rate in the center of the cell is different from the pumping rate calculated before the cell. The pumping rate $R$ in the Eqs.(\ref{psn}), (\ref{spn}), (\ref{fn}) and (\ref{probenoise}) should be replaced by Eq.(\ref{absorption}) evaluated at the location of the probe beam. This modifies the noise limits accordingly.

\section*{Experiments and Results}

\begin{figure*}
\centering
\includegraphics[width=12cm]{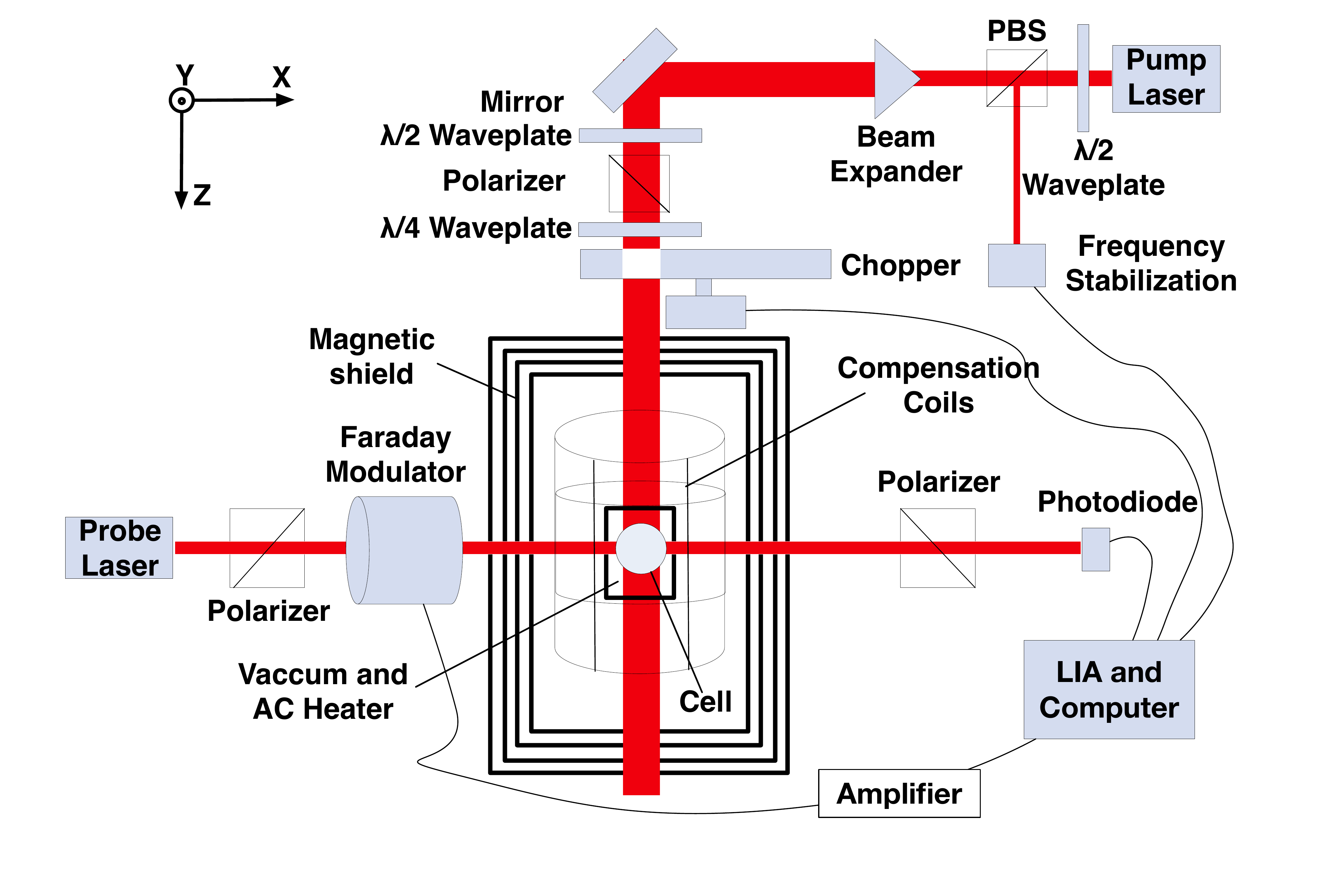}
\caption{\label{experimenal setup} Experimental Setup of a SERF magnetometer.}
\end{figure*}

The experimental setup is shown as Fig. \ref{experimenal setup}. A spherical cell with a diameter of approximately 25 mm is placed in a vacuum chamber containing a drop of potassium, approximately 1600 torr helium buffer gas and 33 torr nitrogen quenching gas for suppressing radiation trapping \cite{molisch1998radiation}. The vacuum chamber is made of G-10 fiberglass, and the cell is heated up to 460 K with an AC heater, which is made of twisted wires to reduce magnetic field. The magnetic-shielding system includes mu-metal magnetic shields and active compensation coils. The shielding factor of the magnetic shield is approximately $10^5$, supplemented by the compensation coils the residual magnetic field at the cell position is smaller than 10 pT. The pump beam propagates along the Z-axis; its wavelength is locked to 770.1 nm (the center of the D1 resonance) to reduce the light shift. The diameter of the pump beam illuminating the cell is approximately 15 mm. 

The probe beam propagates along the X-axis; it is approximately 0.5 nm (250 GHz) detuned to lower frequency from the potassium D2 line. The probe beam is linearly polarized with a Glan-Taylor polarizer. Additionally, a Faraday modulator is used to reduce the 1/f noise at low frequency by modulating the beam polarization with an amplitude of approximately 0.03 rad at a frequency of 5.1 kHz. Then the probe beam passes through another Glan-Taylor polarizer set at 90$^\circ$ to the initial beam polarization direction. A lock-in amplifier (LIA) is used to demodulate the signal from the photodiode.
In order to precisely calibrate the coils of the SERF magnetometer, we applied the synchronous optical pumping technique. By applying a chopper to modulate the pump beam, the magnetometer can work in the Bell-Bloom mode (BB mode) \cite{bell1961optically}.

Calibration of the compensation coils is performed using the applied chopper at frequency ($\omega$) and an additional bias magnetic field in the Y-direction ($B_y$). The response of the BB magnetometer can be written as
\begin{equation}
S_x(\omega)=\frac{R S_0}{4\sqrt{(2\pi \Delta \nu)^2+(\omega-\gamma B_y)^2}}
\label{lweqation},
\end{equation}
where $S_0$ is the polarization in zero magnetic field, $\gamma=\gamma_e/q$, $q$ is the slowing-down factor \cite{appelt1998theory}, which is determined by the polarization of the potassium atoms, $\Delta \nu$ is the magnetic linewidth. 

\begin{figure}
\centering
\includegraphics[width=8cm]{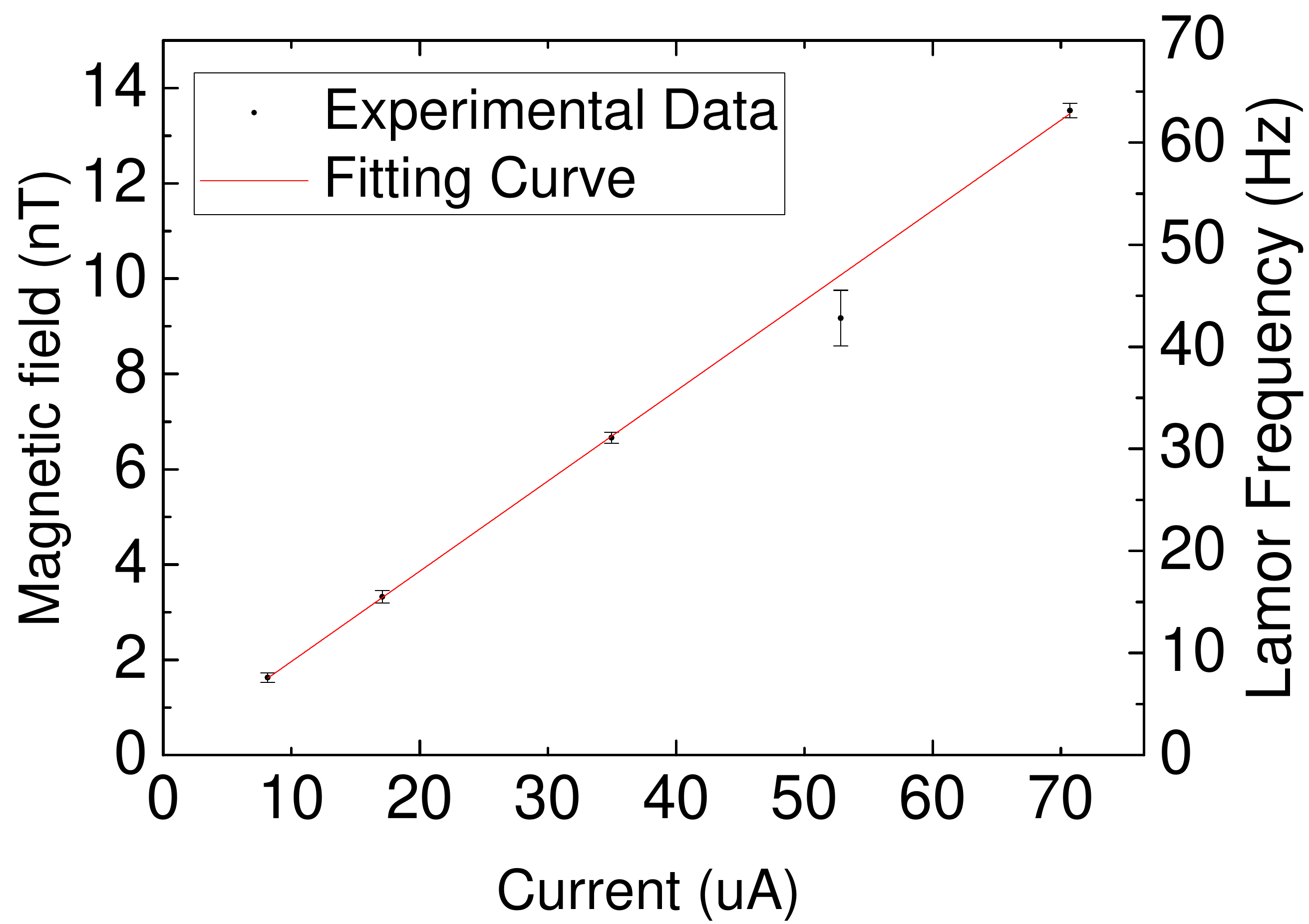}
\caption{\label{Coilc} Linear fit of the magnetic field along the Y-axis as a function of the current applied in the Y coils.}
\end{figure}

In order to keep the nuclear slowing-down factor constant ($\approx$ 6) \cite{appelt1998theory}, the powers of the pump beam and the probe beam are adjusted to small values where the magnetic linewidth is independent of the powers of the pump beam and the probe beam. By applying different bias magnetic fields in the Y-axis, we measured the response of the BB magnetometer. The magnetic field generated by the Y coils can be calculated from the resonant point using Eq.(\ref{lweqation}). The results are shown in Fig. \ref{Coilc}. The measured data near 50 Hz (line frequency in China) has a relatively large error bar, because the lock-in amplifier has a notch filter near the line frequency, which attenuates the response signal. According to the linear fit, we measure the coil calibration constant to be approximately 0.177 nT/$\mu$A.

\begin{figure*}
\centering
\includegraphics[width=12cm]{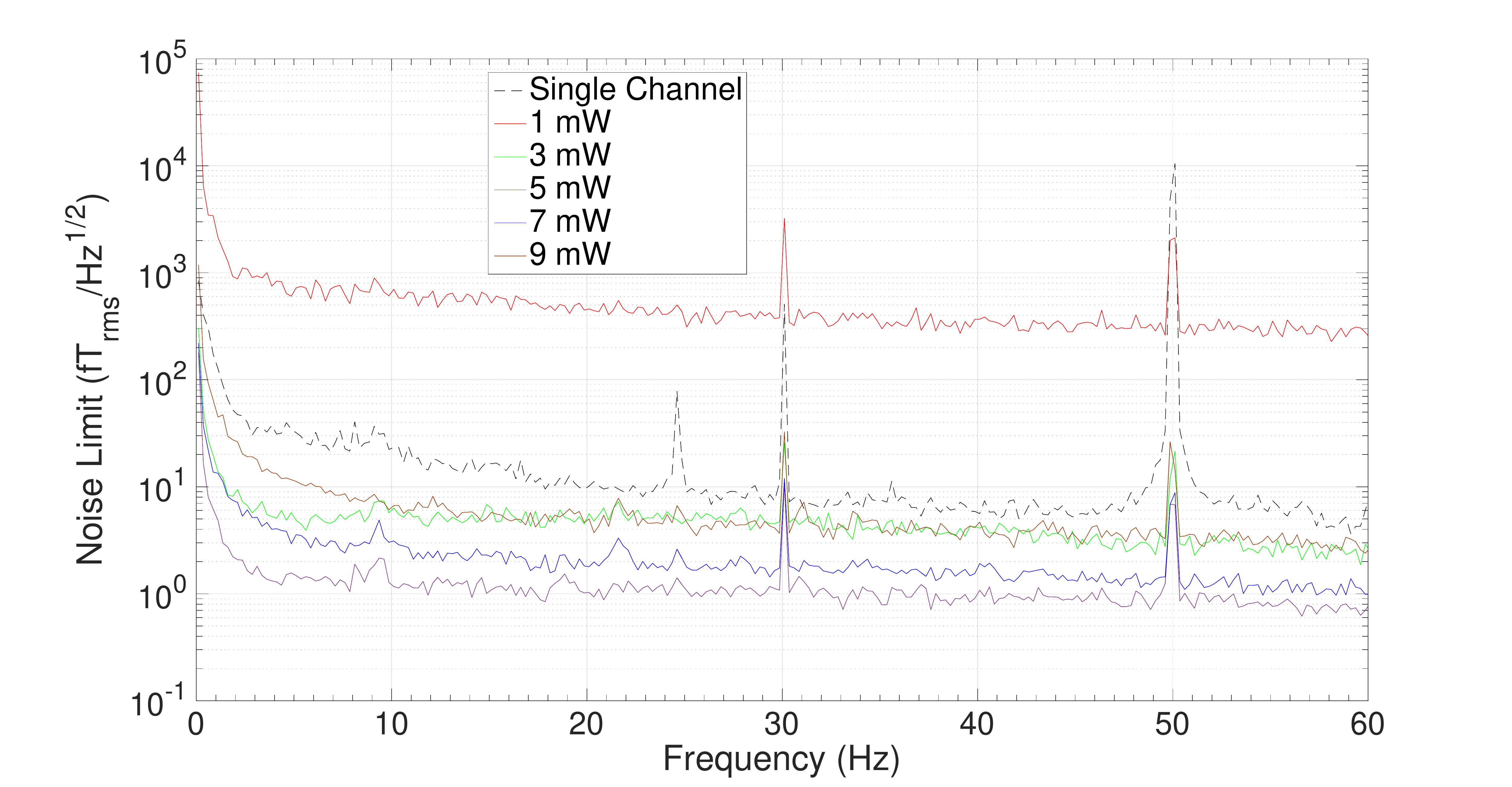}
\caption{\label{fig_fl}The noise limits of the SERF magnetometer with different power of the pump beam. In order to clearly show the curves, only 1 mW, 3 mW, 5 mW, 7 mW and 9 mW experimental results are plotted. The dashed black line is the optimized single-channel (one probe beam, as distinguished from the gradiometer) noise of the SERF magnetometer which is mainly dominated by the magnetic noise of the magnetic shield, it is calibrated by a magnetic field oscillating at 30 Hz with an amplitude approximately 0.25 $\rm{pT_{rms}}$. }
\end{figure*}

After the coil calibration experiments, the chopper is turned off, and the residual magnetic fields are well-compensated to near zero. The power of the probe beam is increased to 0.5 mW and the power of the pump beam increased to 1 mW. A calibration magnetic field oscillating at 30 Hz is applied along the Y-direction to calibrate the response of the magnetometer, whose amplitude is approximately 15.6 $\rm{pT_{rms}}$. Then the sensitivity of the SERF magnetometer at 30 Hz is calibrated. There is Johnson noise of several $\rm{fT/\sqrt{Hz}}$ generated by the mu-metal magnetic shield, which under certain conditions could exceed the intrinsic sensitivity limits (spin projection noise, photon shot noise and technical noise in the optical rotation measurement) of the SERF magnetometer. In order to measure the noise limits of the SERF magnetometer, the pump beam is blocked after the calibration, and the noise floor of the response of the SERF magnetometer is measured and recorded. This procedure enables us to distinguish the noise limit related to quantum noise and technical noise of the probe beam from the Johnson noise of the magnetic shield and pump-beam related noise. Then we increase the power of the pump beam in steps of 1 mW, and repeat the experiments until the power of the pump beam reaches 10 mW. The experimental results are shown in Fig. \ref{fig_fl}, the peaks of the signal responses at 30 Hz are caused by the probe beam's pumping effect and the applied calibration magnetic field \cite{fang2014situ}. For comparison, the magnetic noise limit of the shield is also shown in the figure (single channel), which is the Johnson noise of the magnetic shield, and is approximately 7.5 $\rm{fT/\sqrt{Hz}}$, which matches well with the theoretical prediction. The magnetic noise of a finite length mu-metal magnetic shield can be written as \cite{lee2008calculation}
\begin{equation}
	\delta B_{mag}=\frac{\mu_0}{r}\sqrt{GkT\sigma t_h},
	\label{eq_magnoise}
\end{equation}
where $\mu_0$ is the permeability of vacuum, $G$ is a constant determined by the geometry of the magnetic shield \cite{lee2008calculation}, $k$ is the Boltzmann constant, $T$ is the temperature of the magnetic shield, $\sigma$ is the conductivity of the mu-metal, $t_h$ is the thickness of the innermost magnetic shield, which is approximately 1 mm in the experiment, and $r$ is the radius of the innermost magnetic shield, which is 0.2 m in the experiment.

\begin{figure}
\centering
\includegraphics[width=9cm]{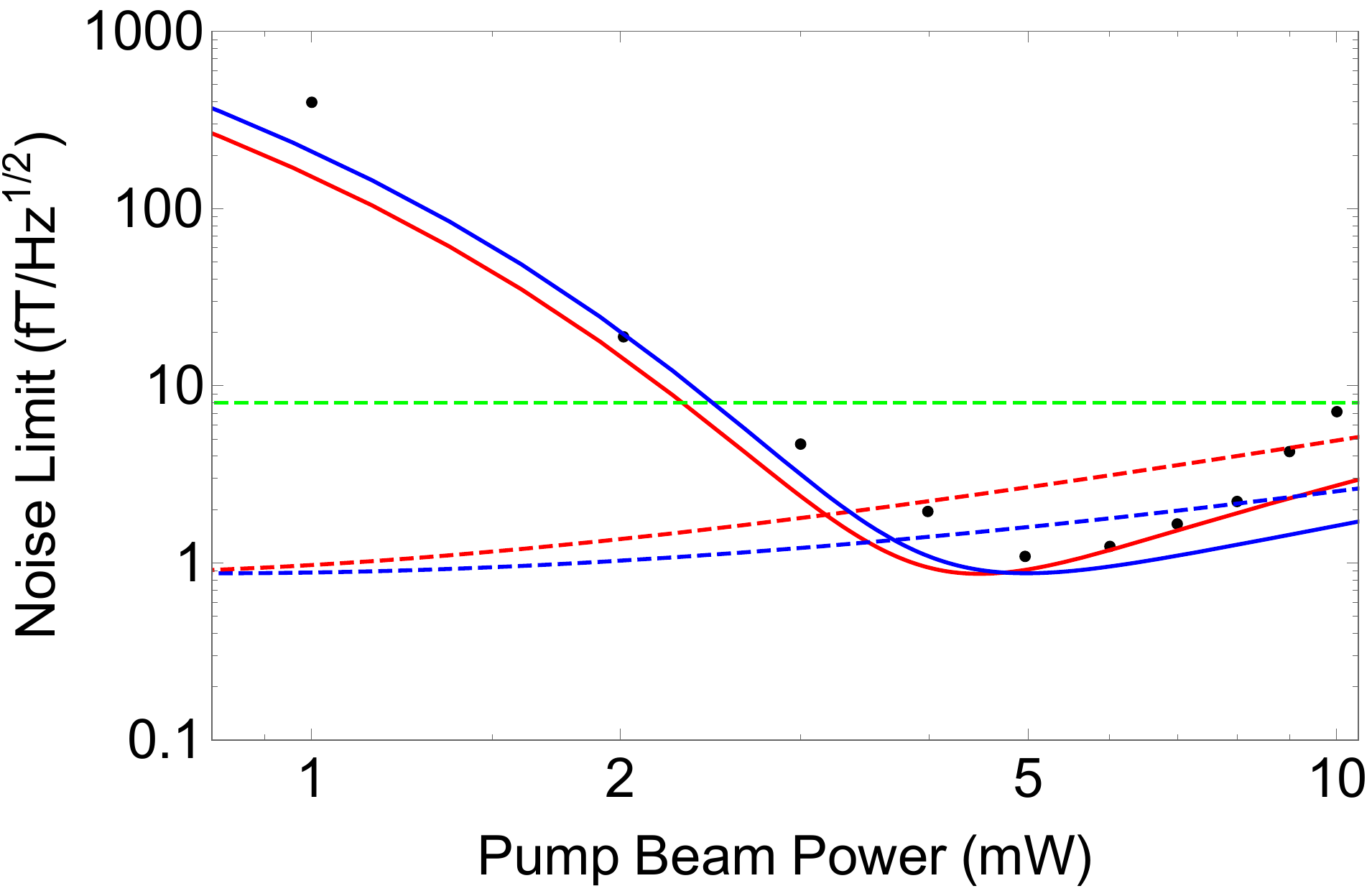}
\caption{\label{experimental noise result} The experimental results and the sensitivity limits based on calculations. The black dots are experimental results, the dashed blue line is the simulation result of the total quantum noise based on Eq.(\ref{fn}), the dashed red line is the technical limit set by the optical rotation sensitivity based on Eq.(\ref{probenoise}), the dashed green line is the magnetic noise of the magnetic shield. The solid red and blue lines are the Eq.(\ref{absorption}) based modified total quantum noise and the technical limit of the optical rotation measurement, respectively. When the pump beam is blocked, the magnetometer is insensitive to the magnetic field, so the noise limit other than the Johnson magnetic noise and the pump beam noise is measured by blocking the pump beam \cite{seltzer2004unshielded}.}
\end{figure}

The frequency response of an undetuned SERF magnetometer is equivalent to a first order low-pass filter with a cutoff frequency equals to $(R+\Gamma_{rel})/q$ \cite{allred2002high}, which means the signal response decreases as the frequency increases. However, there is flicker noise from a magnetic shield below 20 Hz \cite{lee2008calculation,kornack2007low}. Finally, the most sensitive frequency range of a SERF magnetometer is usually between 20 Hz and 40 Hz. In order to demonstrate the relationship between the power of the pump beam and the noise limits, we estimate the noise limits in Fig. \ref{fig_fl}, by calculating the sensitivities around 30 Hz, and plot the results in Fig. \ref{experimental noise result} as black dots. In order to maximize the optical path length of the probe beam propagating through the spherical cell, the probe beam is directed through the center of the cell. The overlapping volume of the probe beam and the pump beam is thus located in the center of the cell. The actual pumping rate of the pump beam should be modified based on Eq.(\ref{absorption}). The noise limits are plotted in Fig. \ref{experimental noise result}. The technical limit is set by the optical rotation sensitivity of approximately $1\times 10^{-7} ~\rm{rad/\sqrt{Hz}}$, which is calibrated by replacing the cell with another known Verdet constant Faraday modulator. According to Fig. \ref{experimental noise result}, when the power of the pump beam is far from sufficient to fully polarize the alkali atoms, the power of the pump beam in the center of the cell attenuates faster than linear. When the power of the pump beam is sufficient to nearly fully polarize the alkali atoms, the pump power attenuates linearly with propagation distance \cite{rochester2002nonlinear}. In the high pump power regime, the technical noise of the optical rotation measurement approaches the intrinsic noise limit of the SERF magnetometer. The experimental results are larger than the theoretical prediction of the technical limits set by the sensitivity of the optical rotation measurement in the high pumping rate region, which could be caused by the non-negligible light shift due to the large pump power and/or the pressure shifts caused by the buffer gas \cite{lwin1978collision}. The modified model for calculating the noise limit of the SERF magnetometer will be helpful in optimizing the power of pump beam, and determining the bottle-neck noise limit of the experimental apparatus.  

The demonstrated noise limit of the apparatus is better than 1 $\rm{fT/\sqrt{Hz}}$, which is still much larger than the fundamental sensitivity of the SERF magnetometer. Comparing with the most sensitive SERF magnetometer mentioned in \cite{dang2010ultra}, our SERF apparatus doesn't have the inner-most low-noise ferrite magnetic shield, and the quantum noise limit of our SERF apparatus could be further improved. This can be achieved by replacing our Faraday modulator and expanding both the pump and probe beams to increase the overlapping volume. (Note that if we increase the sensitivity by expanding both pump and probe beams, it is more accurate to determine the noise limit by averaging the sensitivity instead of applying a single value of z in the Eq.(\ref{absorption}).)

\section{Conclusion}

Modified sensitivity limits of a SERF magnetometer are determined, which consider absorption of the pump beam by the alkali atoms. This absorption modification is demonstrated with a 1 $\mathrm{fT/\sqrt{Hz}}$ SERF magnetometer, where the technical limit set by optical-rotation measurement sensitivity is also identified. SERF magnetometers are currently the most sensitive magnetic sensors in the low-frequency region, whose sensitivity is{} competitive in the SERF-CASPEr experiments. There are several difficulties in using SERF magnetometers in CASPEr, one of which being the limited field range satisfying the condition that the spin-exchange rate exceeds the Larmor precession frequency (the SERF regime). To solve this problem, a superconducting flux transformer is introduced to the SERF-CASPEr experiments which effectively displaces the large sweeping magnetic field away from the SERF magnetometer. Another potential advantage of the flux transformer is the use of low-loss superconducting tunable capacitors to increase the enhancement factor, and corresponding spin precession measurement sensitivity, by working in the tuned mode \cite{seton19974,PhysRevLett.117.141801,savukov2009mri}. Another difficulty only briefly considered above is the thermal isolation required between the secondary coil and the SERF magnetometer, because the cell of the SERF magnetometer needs to be heated to increase the vapor density of the alkali atoms, whereas, the flux transformers need to be cooled to be below critical temperature. However, this can likely be overcome with careful engineering of the experiment. A ferromagnetic needle magnetometer is another potential magnetic sensor that could be applied in future versions of CASPEr experiments, which in principal has a better quantum noise limit and can operate at cryogenic temperatures \cite{kimball2016precessing}. The needle magnetometer does not have the thermal isolation issues, and could benefit from the superconducting transformer (we can easily make the  enhancement factor $\mathrm{k_{FT}}>1$) and the longitudinal detection scheme (extending the limited bandwidth of the needle magnetometer).”

\section{Acknowledgments}
The authors wish to thank Giuseppe Ruoso and Surjeet Rajendran for the useful comments. This project has received funding from the European Research Council (ERC) under the European
Union’s Horizon 2020 research and innovation programme (grant agreement No 695405). We acknowledge the support of the Simons and Heising-Simons Foundations, the DFG Reinhart Koselleck project, the National Science Foundation of the USA under grant PHY-1707875, and the National Science Foundation of China under Grant No. 61227902.

\section*{Appendix: Longitudinal Detection Scheme}
The setup of the SERF-CASPEr-LOD is similar to the SERF-CASPEr setup (as shown in Fig. \ref{Fig_setup_casper-e}), except an additional oscillating field $B_m\mathrm{cos}(\omega_m t)$ is applied perpendicular to ${B_0}$ and the pickup coil is oriented along the ${B_0}$ axis. $\omega_a$ and $\omega_m$ are within the linewidth ($1/T_2$) of the Larmor resonance of the magnetized ferroelectric samples. The primary coil, now oriented along the ${B_0}$ direction, picks up the time-varying magnetization of the sample which can be written as \cite{1742-6596-718-4-042051,Augustine2002111}
\begin{equation}
\begin{aligned}
\Delta M_z=M_z-M_0&\approx\frac{M_0}{4}\gamma_{Pb}^2T_1T_2 B_m B_a \mathrm{cos}[(\omega_m-\omega_a)t]
\\&=n_{Pb}p\mu \gamma_{Pb}T_2B_a\cos[(\omega_m-\omega_a)t]\times \gamma_{Pb}B_mT_1/4
\label{Eq_extand_bandwidth},
\end{aligned}
\end{equation}
where $M_z$ is the longitudinal magnetization of ferroelectric sample, $M_0$ is the static magnetization and $B_m \ll 1/(\gamma_{Pb}\sqrt{T_1T_2})$ to prevent saturation, for example, for $T_2=1 ~\mathrm{s}$, $B_m \ll 0.3 ~\mathrm{nT}$, for $T_2=1 ~\mathrm{ms}$, $B_m \ll 10 ~\mathrm{nT}$. In order to simplify the following calculations, we assume an appropriate amplitude of the oscillating magnetic field $B_m$ to let $\gamma_{Pb}B_mT_1/4=1$, for $T_1\approx$ 1 hr, $B_m \approx$ 20 pT. 

One advantage of applying this strategy in the SERF-CASPEr is that we can keep the SERF magnetometer working in the frequency region corresponding to the optimum sensitivity by tuning the frequency of the oscillating magnetic field $\omega_m$ so that the frequency of the oscillating magnetization ($\omega_m-\omega_a$) is at the optimum. However, the technical noise should be carefully considered when using the LOD scheme. The magnetic noise of the leading field (${B_0}$) will directly couple through the flux transformer and contribute to the noise measured by the SERF magnetometer; the state-of-the-art superconducting magnet system mentioned in \cite{van1999ultrastable} realizes a stability of 17 ppt/hour, which means that for a 10 T leading field from the superconducting magnet, the low-frequency drift of the magnetic field is approximately 170 pT/hour. If the spectrum of the leading magnetic field noise is concentrated mostly in very low frequencies, it may be possible to tune the frequency of the oscillating magnetization far enough away from the peak of the magnetic field noise spectrum to enable a sensitive measurement. The spin projection noise produced by the sample can be estimated as \cite{PhysRevA.73.022107,budker2014proposal}
\begin{equation}
\begin{aligned}
B_{spin} &\approx \mu_0\mu \sqrt{\frac{n_{Pb}}{V_{Pb}}\int_{\omega_0+\delta f-\frac{1}{2\pi T_b}}^{\omega_0+\delta f+\frac{1}{2\pi T_b}} \frac{1}{8} \frac{T_2}{1+T_2^2(\omega_m-\omega_0)^2}  d\omega_p}
\\&= \mu_0\mu \sqrt{\frac{n_{Pb}}{8V_{Pb}T_2} } 
\\& \sqrt{\mathrm{ln} \left[1+T_2^2 \left( \delta_f+\frac{1}{2\pi T_b} \right) \right] -\mathrm{ln} \left[1+T_2^2\left(\delta_f-\frac{1}{2\pi T_b}  \right) \right]}
\label{Eq_spn_sample},
\end{aligned}
\end{equation}
where $V_{Pb}$ is the volume of the sample, $\delta f$ is the offset of the center of the axion signal from the Larmor frequency, here $\delta f= 100 ~\mathrm{Hz}$ (half bandwidth of the SERF magnetometer). For low frequencies (masses) the axion coherence is sufficiently long such that $T_2$ limits ``signal bandwidth time''. When $V_{Pb} \approx 785 ~\mathrm{cm^3}$, here we assume $T_2=1~\mathrm{s}$ leading to $B_{spin} \approx 0.2~\mathrm{aT/\sqrt{Hz}}$. To make sure the technical noise would not surpass the spin-projection noise, the relative amplitude noise
of the pump field should be smaller than $10^{-8}/\mathrm{\sqrt{Hz}}$.

Practically, the noise of the SERF magnetometer is far larger than the spin-projection noise of the ferroelectric sample. If the pump field has a white noise $B_{mn}$, and in order to make sure the magnetic noise would not surpass the sensitivity of the SERF magnetometer, which is assumed to be 50 $\rm{aT/\sqrt{Hz}}$,
\begin{equation}
B_{mn}<\frac{\delta B_{SERF}}{k_{FT} \mu_0 n_{Pb} p \mu \gamma_{Pb} T_2}
\label{Eq_bmn}.
\end{equation}
For Phase 1, $p = 0.001$ and $T_2 = 5 ~\rm{ms}$, $B_{mn} < 40~\mathrm{aT/\sqrt{Hz}}$. For phase 2, $p = 1$ and $T_2 = 1 ~\rm{s}$, $B_{mn} < 0.04~\mathrm{aT/\sqrt{Hz}}$. And $B_m=20~\mathrm{pT}$, so requirement of the amplitude noise of pump field, for Phase 1 is approximately $4\times10^{-6} ~\mathrm{/\sqrt{Hz}}$, for Phase 2 is approximately  $4\times10^{-9} ~\mathrm{/\sqrt{Hz}}$. The sensitivity projection plot of SERF-CASPEr-LOD Phase 1 is shown in Fig. \ref{Fig_Mag2CASPEr}, the measurement time is reduced to 1 hour, because when $T_2=5$ ms it will take approximately 1 year continuously data acquiring to sweep the axion mass up to 1 MHz. For SERF-CASPEr-LOD Phase 2, the measurement time of a single frequency point is reduced to 18 s, and it will take approximately 1 year continuously data acquiring to sweep the axion mass up to 1 MHz.

\begin{figure*}
\centering
\includegraphics[width=10cm]{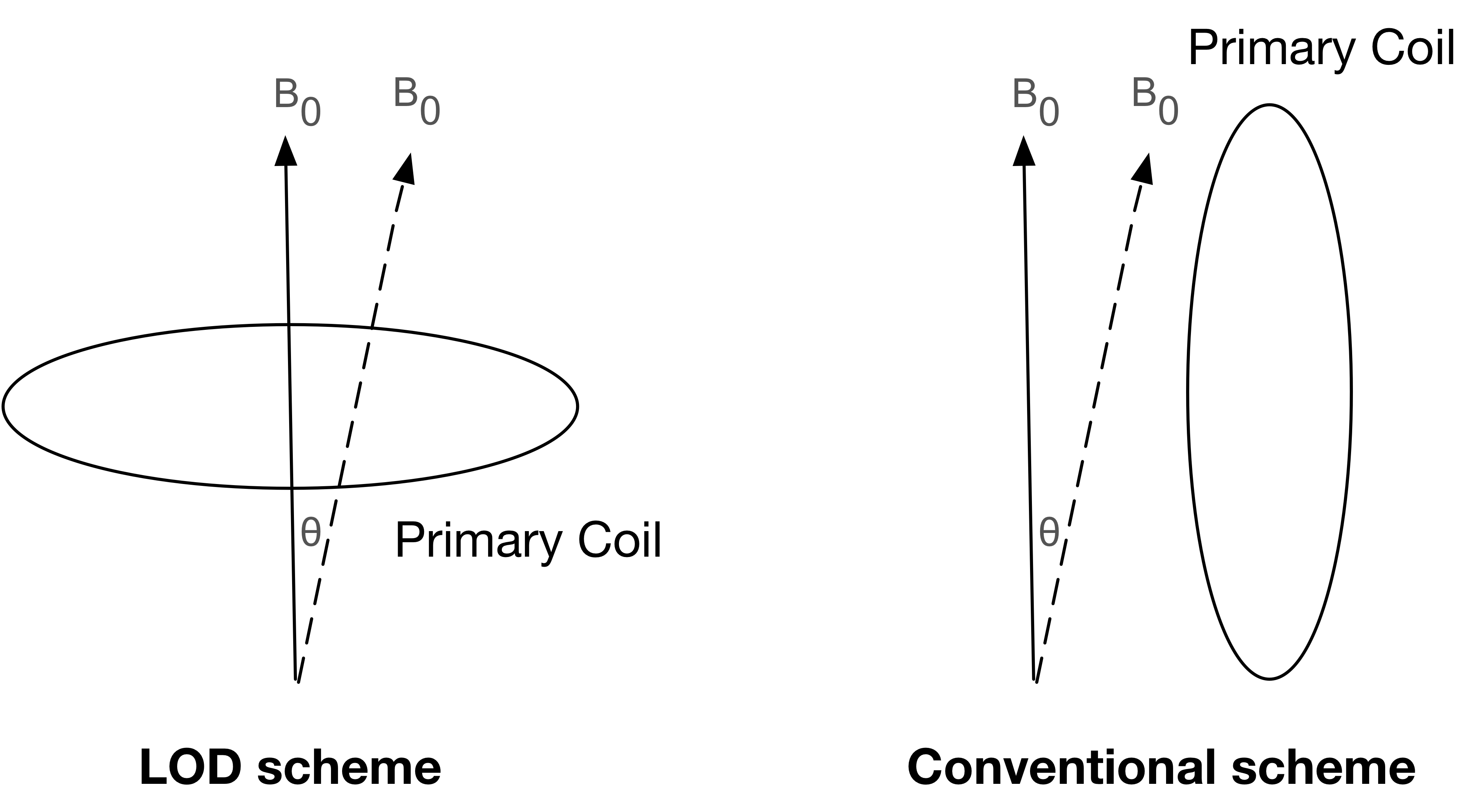}
\caption{Leading field ${B_0}$ projection in the different schemes.}
\label{Fig_vib_coil}
\end{figure*}

We assumed the tilt angle caused by vibration is $\theta \ll 1$. As shown in Fig. \ref{Fig_vib_coil}, in the conventional scheme, the vibrational noise of the leading field ${B_0}$ picked up by the primary coil is 
\begin{equation}
B_{vnc}=B_0\sin(\theta)\approx \theta B_0.
\label{Eq_b_vnc}
\end{equation}
In the LOD scheme, the vibrational noise of the leading field ${B_0}$ picked up by the primary coil is
\begin{equation}
B_{vnl}=B_0\cos(\theta)-B_0=-2B_0\sin^2(\theta/2) \approx -\theta^2 B_0/2.
\label{Eq_b_vnl}
\end{equation}
According to Eqs. (\ref{Eq_b_vnc}) and (\ref{Eq_b_vnl}), the vibrational noise is quadratically suppressed in the LOD scheme which may become a distinct advantage in the event that the sensitivity of CASPEr is limited by vibrational noise.

\bibliography{noiselimit.bib}
\end{document}